\documentclass[sigconf]{acmart}
\usepackage{makecell}

\AtBeginDocument{%
  \providecommand\BibTeX{{%
    \normalfont B\kern-0.5em{\scshape i\kern-0.25em b}\kern-0.8em\TeX}}}

\setcopyright{acmcopyright}
\copyrightyear{2022}
\acmYear{2022}
\acmDOI{XXXXXXX.XXXXXXX}

\begin{document}

\title{Attack Techniques and Threat Identification for Vulnerabilities}
% \subtitle{Industry}

\author{Constantin Adam}
%\email{cmadam@us.ibm.com}
\author{Muhammed Fatih Bulut}
%\email{mfbulut@us.ibm.com}
\author{Daby Sow}
\email{{cmadam, mfbulut, sowdaby}@us.ibm.com}
\affiliation{%
  \institution{IBM T.J. Watson Research Center}
  \city{Yorktown Heights}
  \state{NY}
  \country{USA}
}

\author{Steven Ocepek}
%\email{Steve.Ocepek@ibm.com}
\author{Chris Bedell}
\email{{steve.ocepek, christopher.bedell}@ibm.com}
\affiliation{%
  \institution{IBM Security X-Force Red}
  \streetaddress{}
  \city{}
  \state{}
  \country{USA}}

\author{Lilian Ngweta}
\email{ngwetl@rpi.edu}
\affiliation{%
  \institution{Rensselaer Polytechnic Institute}
  \city{Troy}
  \state{NY}
  \country{USA}
}

\renewcommand{\shortauthors}{Adam and Bulut, et al.}

\begin{abstract}

% To secure organizations' computing environments, it is important not only to understand the technical descriptions of the vulnerabilities, but also to gain insights into attackers' perspectives.  In this work, we use machine learning and natural language processing techniques to map vulnerabilities to attack techniques and threat actors. This is a novel approach that is bridging the gap between the vulnerability descriptions, usually written from the developers' perspective, and the security reports describing real-world exploits and weaponization of those vulnerabilities, often written from the attackers' perspective. This work provides new security intelligence, by predicting which attack techniques are most likely to be used to exploit a given vulnerability and which threat actors are most likely to conduct the exploitation. 

% Abstract from Steve
Modern organizations struggle with what is often considered an insurmountable number of vulnerabilities that are discovered and reported by their network and application vulnerability scanners. Therefore, prioritization and focus become critical, to spend their limited time on the highest risk vulnerabilities. In doing this, it is important for these organizations not only to understand the technical descriptions of the vulnerabilities, but also to gain insights into attackers' perspectives. Due to the sheer size of the data set, this needs to be done at scale to be effective. In this work, we use machine learning and natural language processing techniques, as well as several publicly available data sets to provide an explainable mapping of vulnerabilities to attack techniques and threat actors. This is a novel approach that relies exclusively on publicly available data sets and is bridging the gap between the vocabularies of vulnerability descriptions, usually written from the developers' perspective, and those of the security reports describing real-world exploits and weaponization of those vulnerabilities, often written from the attackers' perspective. This work provides new security intelligence, by predicting which attack techniques are most likely to be used to exploit a given vulnerability and which threat actors are most likely to conduct the exploitation.  Lack of labeled data and different vocabularies make mapping vulnerabilities to attack techniques at scale a challenging problem that cannot be addressed easily using supervised or unsupervised (similarity search) learning techniques. To solve this problem, we first map the vulnerabilities to a standard set of common weaknesses, and then common weaknesses to the attack techniques. This approach yields a Mean Reciprocal Rank (MRR) of $0.95$, an accuracy comparable with those reported for state-of-the-art systems. Our solution has been deployed to IBM Security X-Force Red Vulnerability Management Services, and in production since 2021. The solution helps security practitioners to assist customers to manage and prioritize their vulnerabilities, providing them with an explainable mapping of vulnerabilities to attack techniques and threat actors. As a case study, we present the analysis performed by our solution for the \emph{log4j} vulnerabilities, one of the most serious security breaches that unfolded at the end of 2021.

%After implementing and deploying the solution, we discuss the analysis of the log4j vulnerabilities, one of the most serious security breaches, that unfolded at the end of the last year. 

\end{abstract}




\maketitle

\section{Introduction}

%Hybrid cloud has emerged as the new normal for traditional enterprises looking for ways to leverage the cloud for their digital transformation.  This is driven by the need to create new revenue streams, deliver superior user experience, and reduce capital and operational expenditure.  Today, 94\% of enterprises have a mix of cloud models - public, dedicated, private, and hybrid and 67\% of them are using multiple public clouds.

%Hybrid environments empower the enterprise to access 'unlimited' resources, services, and technologies to accelerate their innovation and their transformation. However, continuing to operate in a secure and compliant way - and adhering to the regulatory requirements across a growing number of providers, tools, and technologies - is a very complex and daunting task.

%To make the situation more difficult, 
Cyberattacks, data breaches and malware are on the rise. On average, over 40 vulnerabilities are reported daily based on the statistics from the National Vulnerability Database (NVD) \cite{nvd-stats}.  Within enterprises, the amount of work to manage security and compliance has become overwhelming. To address these challenges, risk assessment enables analysis of the risk of the detected vulnerabilities, considering a comprehensive set of risk factors: exposure, weaponization, and impact of the deviations discovered through policy execution.

The aim of this work is to associate each vulnerability with one or more attack techniques, and with the threat actors (attackers) known to have used those techniques in the past.  Mapping vulnerabilities to threat actors and attacker techniques enables three fundamental capabilities for vulnerability and risk management. First, it assists security analysts to model potential attacks and then prove or disapprove the attacks effectiveness against the organization through well designed pen testing. Second, it provides the capability to alert a client, given a set of vulnerabilities that were discovered in their system, on the likelihood that these vulnerabilities could be exploited by a threat actor they are concerned about. Third, when new vulnerabilities are published, it generates a list of threat actors that are the most likely to weaponize them and brings the new vulnerability to the attention of the clients that are most likely to be impacted by the threat actors in the list.

There are two primary challenges for mapping vulnerabilities to attack techniques and threat actors. Lack of labelled data makes it hard to train a model using supervised learning techniques. Also unsupervised techniques such as similarity search with cosine similarity, is inadequate for the task, because different vocabularies are used for describing vulnerabilities and the attack techniques.

%By parsing the description of the MITRE ATT\&CK techniques, we found 45 vulnerabilities that are referenced under various techniques description, and can therefore be associated with those techniques.  This data is clearly not sufficient to train a supervised machine learning model for over $160,000$ vulnerabilities reported so far in NVD. Moreover, unsupervised machine learning techniques such as similarity learning, based on text similarity between vulnerabilities and attack techniques, do not perform well enough to be used in production settings.

% To bridge these gaps, we break the process of mapping vulnerabilities to attack techniques in two steps.  First, we map vulnerabilities to Common Weaknesses (a list of software and hardware weakness types produced by a MITRE-led community effort) \note{This is the first time we are talking about MITRE, either reference or an explanation needed. This is an AI community, they probably are not familiar with MITRE}, and second, we map Common Weaknesses to attack techniques.  

To bridge these gaps, we break the process of mapping vulnerabilities to attack techniques in two steps: first, we map vulnerabilities to Common Weaknesses, and second, we map Common Weaknesses to attack techniques.  Common Weaknesses~\cite{cwe} are flaws, faults, bugs, or other errors in software or hardware implementation, code, design, or architecture that, if left unaddressed, could result in systems, networks, or hardware being vulnerable to attack.  MITRE~\cite{mitre} has led a community effort that produced a list of common software and hardware weakness types that have security ramifications.

As many vulnerabilities are already mapped to valid Common Weaknesses (about $70\%$ of the existing vulnerabilities) by experts, we can utilize this ground truth data and train a machine learning model that can provide the mappings for all vulnerabilities. This two-step approach also reduces the dimensionality of the problem.  There are about $900$ common weakness types~\cite{cwe}, that change infrequently, and are easier to map to attack techniques than the $170,000$ recorded vulnerabilities, as of June 2022.  As the common weaknesses constitute a hierarchical structure, the problem is further simplified, as the parents in the common weakness tree are already mapped to all the attack techniques of the children.  Once the vulnerabilities are mapped to the attack techniques, the Cloud Security Operations Center can use public data from~\cite{mitre-attack} to map the attack techniques to threat actors

We have implemented our solution and deployed a vulnerability risk analyzer that associates each vulnerability with one or several attack (weaponization) techniques, and with the threat actors that have used those techniques in the past.  The analyzer then highlights the vulnerabilities that are most likely to be exploited in the future by the threat actors that an organization is concerned about. After completing deployment, we have reviewed the results that our analyzer produced when multiple threat actors began exploiting the log4j family of vulnerabilities, one of the most serious security breaches that unfolded in December 2021, and that affected much of the Internet.

Our contributions are as follows:
\begin{itemize}
 \item Two-step approach, relying exclusively on public data sets, to map vulnerabilities to attack techniques by bridging the gap between attackers, their techniques and vulnerabilities,
 \item Hierarchical machine learning based approach to map vulnerabilities to CWEs, and experiments to validate the accuracy of it,
 \item Evaluation and comparison of an unsupervised machine learning (similarity learning) and supervised machine learning (two-step approach) approaches, and
 \item Discussion of how the mapping of vulnerabilities to attack techniques has analyzed the log4j security breach, and the lessons learned.
\end{itemize}

The rest of the paper is organized as follows. Section~\ref{sec:background} describes the background for this work.  Section~\ref{sec:method} explains the system design and approach to solve the problem. Section~\ref{sec:eval} presents the performance evaluation. Section~\ref{sec:deployment} covers the deployment of our methodology, discusses the analysis results for the log4j security breach, and describes the application use cases.  Section~\ref{sec:related} reviews the related work and finally we conclude with Section~\ref{sec:ftr}.
\section{Background}
\label{sec:background}
\begin{table*}
  \caption{Examples of text description in natural language for a vulnerability and attack technique}
  \label{table:example-texts}
  \begin{tabular}{ccc}
    \toprule
    Type &Identifier & Description\\
    \midrule
    Vulnerability & CVE-2018-1977& \makecell{IBM DB2 for Linux, UNIX and Windows 11.1 (includes DB2 Connect Server) \\contains  a denial of service vulnerability. A remote, authenticated DB2 user \\could exploit this vulnerability by issuing a specially-crafted \\SELECT statement with TRUNCATE function.} \\
    Attack Technique & T1531 & \makecell{
    Adversaries may interrupt availability of system and network resources by \\inhibiting access to accounts utilized by legitimate users. Accounts may be \\deleted, locked, or manipulated (ex: changed credentials) to remove \\access to accounts.} \\
    \bottomrule
  \end{tabular}
\end{table*}
% \begin{table*} [t]
%\scriptsize
% \begin{center}
%   \caption{Examples of text description in natural language for a vulnerability and attack technique}
%   %\vspace{-1mm}
%   \label{table:example_texts}
%   %\begin{tabular}{ | m{1cm} | m{10.0cm} | l | l | }
%   \begin{tabular}{ | m{1.7cm} | m{2.4cm} | m{12cm} | }
%     \hline
%     {\bf Type} & {\bf Identifier} & {\bf Description} \\ \hline \hline
%     Vulnerability & CVE-2018-1977 & IBM DB2 for Linux, UNIX and Windows 11.1 (includes DB2 Connect Server) contains a denial of service vulnerability. A remote, authenticated DB2 user could exploit this vulnerability by issuing a specially-crafted SELECT statement with TRUNCATE function. \\ \hline
%     Attack Technique & T1531 & Adversaries may interrupt availability of system and network resources by inhibiting access to accounts utilized by legitimate users. Accounts may be deleted, locked, or manipulated (ex: changed credentials) to remove access to accounts. \\ \hline
%   \end{tabular}
  % \vspace{-7mm}
% \end{center}
% \end{table*}

\subsection{Vulnerabilities}

Software vulnerabilities are flaws, weaknesses that are present in code. Vulnerabilities are exploited by attackers to conduct various type of attacks such as ransomware, phishing, malware infection or data leakage. Each vulnerability is uniquely identified with a specific id (CVE - Common Vulnerability Exposure) and has a description, written in natural language, in a free form text format.  The description details the specifics of product(s) that are affected, and the exploitation thereof, from a developer's perspective. CVEs can be enriched with additional information such as Common Vulnerability Scoring System (CVSS) assessments, affected products and Common Weakness Enumerations (CWEs).

% \begin{figure}[h]
%   \centering
%   \includegraphics[width=\linewidth]{images/cve_trends.pdf}
%   \caption{Number of CVEs per year since 2020}
%   \label{fig:cve_trends}
% \end{figure}

% In recent years, the number of reported vulnerabilities has been steadily increasing. Figure~\ref{fig:cve_trends} shows the number of vulnerabilities per year where spikes in recent years have been prevalent. There are various factors that cause this proliferation including increase in software products, open source software, move to public cloud environments and also lucrative successful attack campaigns (such as ransomware). %There are various work that try to understand the vulnerabilities and their impacts \cite{}.

\subsection{MITRE ATT\&CK Framework}

To exploit a vulnerability, an attacker must have at least one applicable tool or technique that can connect to a weakness existing in the system. The MITRE ATT\&CK framework is a curated knowledge base and model for cyber adversary behavior, reflecting the various phases of an adversary's attack life cycle and the platforms they are known to target~\cite{mitre-attack}. 

MITRE ATT\&CK framework is organized in a hierarchical structure. There are top techniques and underneath each, there are sub-techniques. For example, in the ``Credential Access'' category there 15 techniques listed and one of them is ``Brute Force''. There are 4 children of ``Brute Force'', namely: ``Password Guessing'', ``Password Cracking'', ``Password Spraying'' and ``Credential Stuffing''. As one traverses this hierarchy from a technique to sub-techniques, the information provided becomes more specific and more detailed.

\subsection{Challenges Mapping Vulnerabilities to Attack Techniques}
Table~\ref{table:example-texts} shows an example of a vulnerability mapped to an attack technique (\emph{T1531}). Since there is not much training data that ties vulnerabilities to ATT\&CK techniques, direct mapping from CVE to ATT\&CK techniques can be made possible via unsupervised approaches such as text similarity. However, traditional text similarity techniques such as cosine similarity perform poorly for the examples in Table~\ref{table:example-texts}.  Even when utilizing an embedding technique (word2vec) to capture the semantic aspect of the text, resulting a cosine similarity score of $0.247$, where 0 means not similar at all, 1 means identical. Besides, from a broader perspective, if we look at the top 20 words that are used in the descriptions of vulnerabilities and of MITRE ATT\&CK techniques' descriptions separately, only 4 out of 20 are common and the remaining 16 are different. Hence, using unsupervised text similarity-based approach would have limits to achieve our objective of mapping CVEs to MITRE ATT\&CK techniques.

\subsection{Common Weakness Enumeration (CWE)}
\label{subsec:cwe}
A vulnerability is a weakness which can be exploited by a threat actor, such as an attacker, to cross privilege boundaries within a computer system. To connect vulnerabilities to techniques, we use the Common Weakness Enumeration (CWE)~\cite{cwe}.  CWE is a community-developed list of software and hardware weakness types. It serves as a common language, a measuring stick for security tools, and as a baseline for weakness identification, mitigation, and prevention efforts.

Like MITRE ATT\&CK techniques, CWE is organized in a hierarchical structure. The parents represent generic weaknesses, and their children subsequently refine the weakness description, apply to specific platforms, or scenarios.  For example, CWE-707 refers to a generic category of weaknesses named ``Improper Neutralization'', its child CWE-74 covers injection of data - ``Improper Neutralization of Special Elements in Output Used by a Downstream Component ('Injection')'', the next child CWE-79 covers cross-site scripting, a particular form of injection (``Improper Neutralization of Input During Web Page Generation ('Cross-site Scripting')''), and finally the leaf common weakness CWE-80 refers to a particular way of doing cross-site scripting ``Improper Neutralization of Script-Related HTML Tags in a Web Page (Basic XSS)''

CWE establishes an abstraction of vulnerabilities into a hierarchical limited set of weakness types. Moreover, each weakness type constitutes an attack surface that ties to attack techniques. There has been a MITRE led effort~\cite{capec, capec-cwe} to connect CWEs to Common Attack Pattern Enumeration and Classification (CAPEC).  Similarly, a mapping of CWE to MITRE ATT\&CK techniques can be established to reduce to mapping CVEs to MITRE ATT\&CK techniques to a problem of mapping CVEs to CWEs, in which there are many training samples that can be used as a part of supervised machine learning approaches.  Unfortunately, we cannot map CWEs to attack techniques through CAPEC attacks, because the data is not complete.  Out of the 891 CWEs, only 120 are mapped to 339 out of the 530 CAPEC attacks.  Furthermore, only 79 CAPEC attacks are mapped to 89 attack techniques.  So, in total, only 41 CWEs are mapped to 89 Mitre Attack techniques.

% \subsection{Integration with Security Architecture}
% Figure~\ref{fig:arch} shows the integration of our classification methodology in a security architecture.  

% \subsection{Data Feeds}
% The system takes data feeds from three sources: the National Vulnerability Database (vulnerabilities descriptions and their mapping to common weaknesses), the MITRE Common Weaknesses Enumeration (common weaknesses description and their mapping to vulnerabilities), and the MITRE ATT\&CK website (descriptions of the attack techniques). 

% \subsubsection{Use Cases}
% We are using a machine learning model to map vulnerabilities to common weaknesses.  A security SME can visualize the mappings of vulnerabilities to common weaknesses returned by the model and refine / correct them, using active learning to improve the mapping accuracy.  The mappings are stored in a database that is accessible through an API to other services, such as risk and vulnerability management, penetration testing, or security event management.

% \begin{figure}[h]
%   \centering
%   \includegraphics[width=\linewidth]{images/arch.pdf}
%   \caption{Methodology integration in a security framework}
%   \label{fig:arch}
% \end{figure}
\section{System Design}
\label{sec:method}

Our workflow of mapping vulnerabilities to attack techniques consists of two steps: mapping vulnerabilities to common weaknesses and mapping common weaknesses to attack techniques.  In this section we describe these steps in more detail.

\subsection{Mapping Vulnerabilities to Common Weaknesses}

We use a methodology for mapping vulnerabilities to common weaknesses like the one described in~\cite{threatzoom}. First, we preprocess the text, by cleaning it up (converting to lower case, filtering the stop words, and removing punctuation or special characters), stemming words, and applying synonym word coding.  Synonym word coding consists of replacing certain word sequences that are used frequently to describe vulnerabilities of a specific weakness type with code words.  MITRE provides a section for a portion of CWEs called ``Alternative Terms'' in which it provides the abbreviations or other commonly used advanced terms for that particular weakness.  For example, [Buffer Overflow, buffer overrun] is a word vector associated with CWE-119 ('Improper Restriction of Operations within the Bounds of a Memory Buffer').  MITRE also provides a ``CWE Glossary'' which contains more general terminology that is interchangeably used in CWE descriptions to convey similar meanings. Each group of words is represented by a code that is a word from the group, such that any other word in the group will be replaced by this code if found in the text.

Second, we perform feature extraction, where we turn the preprocessed text into a set of numerical vectors.  We experiment with two types of feature extraction.  The first is based on TF-IDF and N-Grams.  The second is based on word embedding using word2vec.

Third, we use a set of neural networks (one single layer neural network with sigmoid activation per common weakness) to associate vulnerabilities with common weaknesses.  As described in Background section, this is a multi-label hierarchical classification problem, as common weaknesses are structured in a hierarchical graph, and a vulnerability can be associated with one or several common weaknesses.  The neural networks use the feature vectors computed in the second step as input and output the CWE classes associated with the vulnerability of interest.  The neural networks are classifiers, and the classification process of a vulnerability to common weaknesses is following the CWE hierarchy.  Any incoming vulnerability is first assigned by a classifier to one of the 10 ``root'' common weaknesses that do not have any ancestors in the CWE hierarchy graph.  We also define a threshold, and the common weaknesses that are assigned to a vulnerability with a score above that threshold are added to the mapping results.  If no score is above the threshold, then we select the root common weakness with the highest score.  This process is then repeated at each level, for the successors of the ``root'' common weaknesses, for their successors, and so on, until we reach the leaves of the CWE hierarchy.

% We have experimented with other multi-label hierarchical classifiers, such as~\cite{DBLP:journals/corr/abs-2010-10151}, that use a single neural network for classification, but then incorporate the hierarchy information in the loss function, by ensuring that the probability of assigning a vulnerability $V$ to any common weakness $W$ is higher than assigning $V$ to any of the successors of $W$ in the hierarchy graph.  But the results were not acceptable.

\subsection{Mapping CWEs to MITRE ATT\&CK Techniques}

%We are using existing mappings to match common weaknesses with CAPEC Attacks (we view the problem of mapping common weaknesses to Mitre Attack Techniques as similar, with the mappings being built by hand and using similarity search and reviewed by an SME).  

Our methodology for mapping CWEs to MITRE ATT\&CK techniques relies on looking up data in several tables, generated from publicly available data sets.  First, as described in Section~\ref{sec:background}, we use~\cite{capec, capec-cwe} to connect CWEs to Common Attack Pattern Enumeration and Classification (CAPEC), and then the CAPEC data to MITRE ATT\&CK techniques.  
In total, we can map 41 CWEs to 89 Mitre Attack techniques using the CAPEC data set.  Next, we parse the MITRE ATT\&CK dataset, and extract all the vulnerabilities (CVEs) that are associated with attack techniques in the techniques description.  Finally, we use a public data set~(\cite{cwe-attack-map}) that maps CWEs to attack techniques.
\section{Experiments}
\label{sec:eval}

\begin{table*}
  \caption{Distribution of vulnerabilities to 10 root common weaknesses}
  \label{table:cve-cwe-distribution}
  \begin{tabular}{ccc}
    \toprule
      CWE & Title & Number of CVEs\\
      \midrule
      CWE-284 & Improper Access Control & 8680 \\
      CWE-435 & Improper Interaction Between Multiple Correctly-Behaving Entities & 83 \\
      CWE-664 & Improper Control of a Resource Through its Lifetime & 38363 \\
      CWE-682 & Incorrect Calculation & 890 \\
      CWE-691 & Insufficient Control Flow Management & 8822 \\
      CWE-693 & Protection Mechanism Failure & 2748 \\ 
      CWE-697 & Incorrect Comparison & 43 \\
      CWE-703 & Improper Check or Handling of Exceptional Conditions & 1046 \\
      CWE-707 & Improper Neutralization & 50074 \\
      CWE-710 & Improper Adherence to Coding Standards & 2072 \\
      \bottomrule
  \end{tabular}
\end{table*}

% \scriptsize
% \begin{table}
%   \footnotesize
%   \centering
%   \caption{Distribution of vulnerabilities to 10 root common weaknesses}
%   \begin{tabular}{m{1.3cm} | m{4.4cm} | m{1.4cm}}
%       \hline\noalign{\smallskip} \textbf{CWE} &
%       \textbf{Title} & \textbf{\# of CVEs}\\
%       \noalign{\smallskip}\hline
%       CWE-284 & Improper Access Control & 8680 \\
%       CWE-435 & Improper Interaction Between Multiple Correctly-Behaving Entities & 83 \\
%       CWE-664 & Improper Control of a Resource Through its Lifetime & 38363 \\
%       CWE-682 & Incorrect Calculation & 890 \\
%       CWE-691 & Insufficient Control Flow Management & 8822 \\
%       CWE-693 & Protection Mechanism Failure & 2748 \\ 
%       CWE-697 & Incorrect Comparison & 43 \\
%       CWE-703 & Improper Check or Handling of Exceptional Conditions & 1046 \\
%       CWE-707 & Improper Neutralization & 50074 \\
%       CWE-710 & Improper Adherence to Coding Standards & 2072 \\
%       \noalign{\smallskip}\hline
%   \end{tabular}
%   \label{table:cve-cwe-distribution}
% \end{table}
% \normalsize

\subsection{Mapping Vulnerabilities to Common Weaknesses}

Our data set contains mappings of $113,000$ vulnerabilities to common weaknesses, out of the total of $170,000$ vulnerabilities registered in the National Vulnerability Database, as of June 2022 (we have filtered out the vulnerabilities mapped to \emph{NVD-CWE-noinfo} and \emph{NVD-CWE-Other}). These vulnerabilities are mapped to $266$ CWEs.  Table~\ref{table:cve-cwe-distribution} shows the number of vulnerabilities mapped to each of the $10$ common weaknesses at the top of the CWE hierarchy. 

For evaluation, the data set is divided into 3 groups: training ($70\%$), validation ($10\%$) and testing ($20\%$).  We analyze the data using the methodology described in~\cite{Sokolova:2009:SAP:1542545.1542682} and calculate precision, recall, and F-score by taking the average of $k$ iterations for the \emph{micro} and \emph{macro} measures. Each calculation was limited to the union of labels that exist within actual and predicted classes.

% \begin{figure}[h]
%   \centering
%   \includegraphics[width=\linewidth]{images/cve-cwe-distribution.pdf}
%   \caption{Distribution of vulnerabilities to 10 root common weaknesses}
%   \label{fig:cve-cwe-distribution}
% \end{figure}

% \begin{equation}
% \label{eq:precision}
% P = \frac{T_p}{T_p + F_p}
% \end{equation}

% \begin{equation}
% \label{eq:recall}
% R = \frac{T_p}{T_p + F_n}
% \end{equation}

% \begin{equation}
% \label{eq:f_score}
% Fscore = \frac{P . R}{P + R}
% \end{equation}

\begin{figure}[h]
  \centering
  \includegraphics[width=\linewidth]{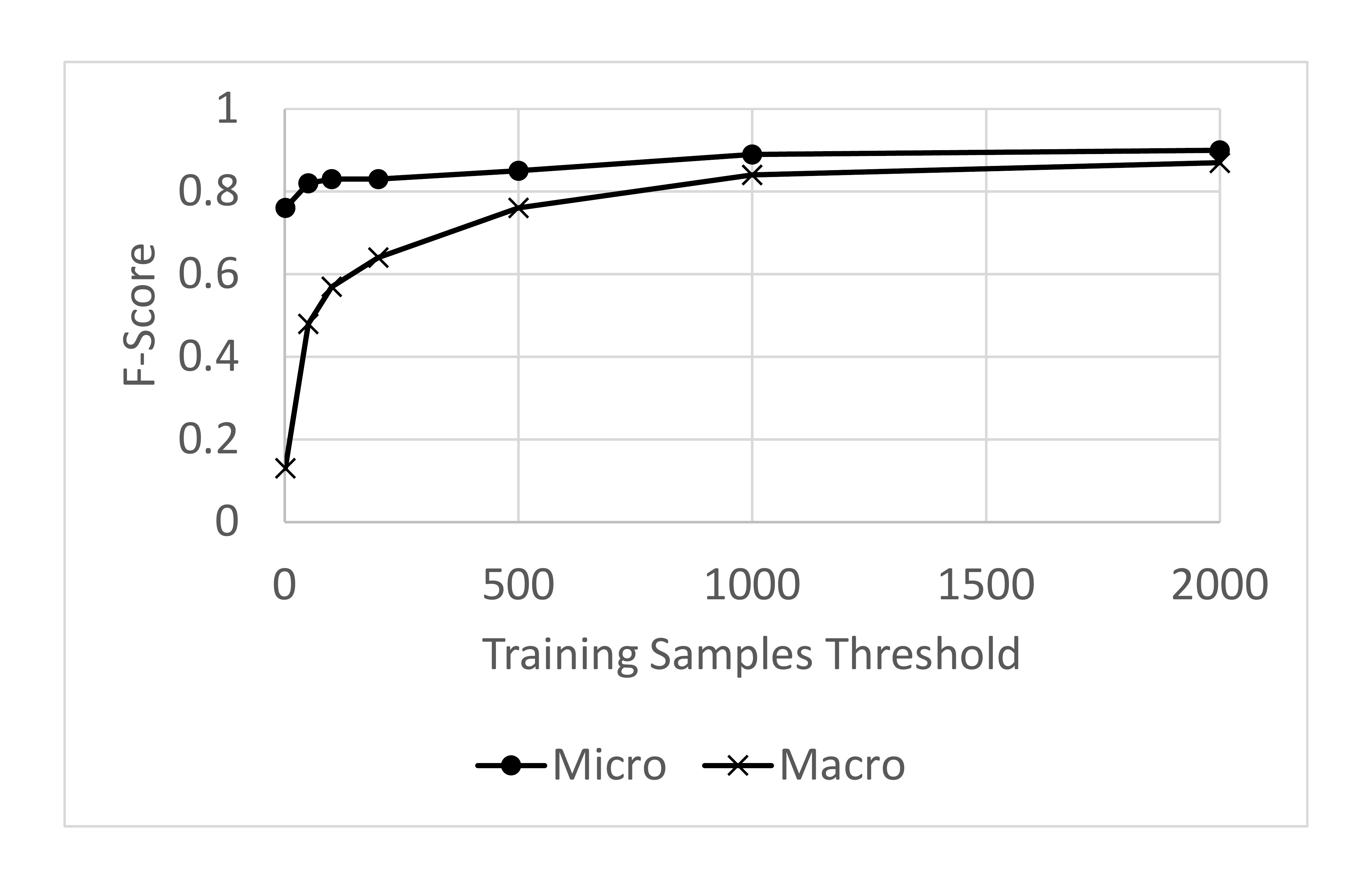}
  % \vspace{-1cm}
  \caption{Micro and macro F-scores for vulnerability to common weakness classification}
  \label{fig:cve-cwe-fscore}
\end{figure}

\begin{figure}[h]
  \centering
  \includegraphics[width=\linewidth]{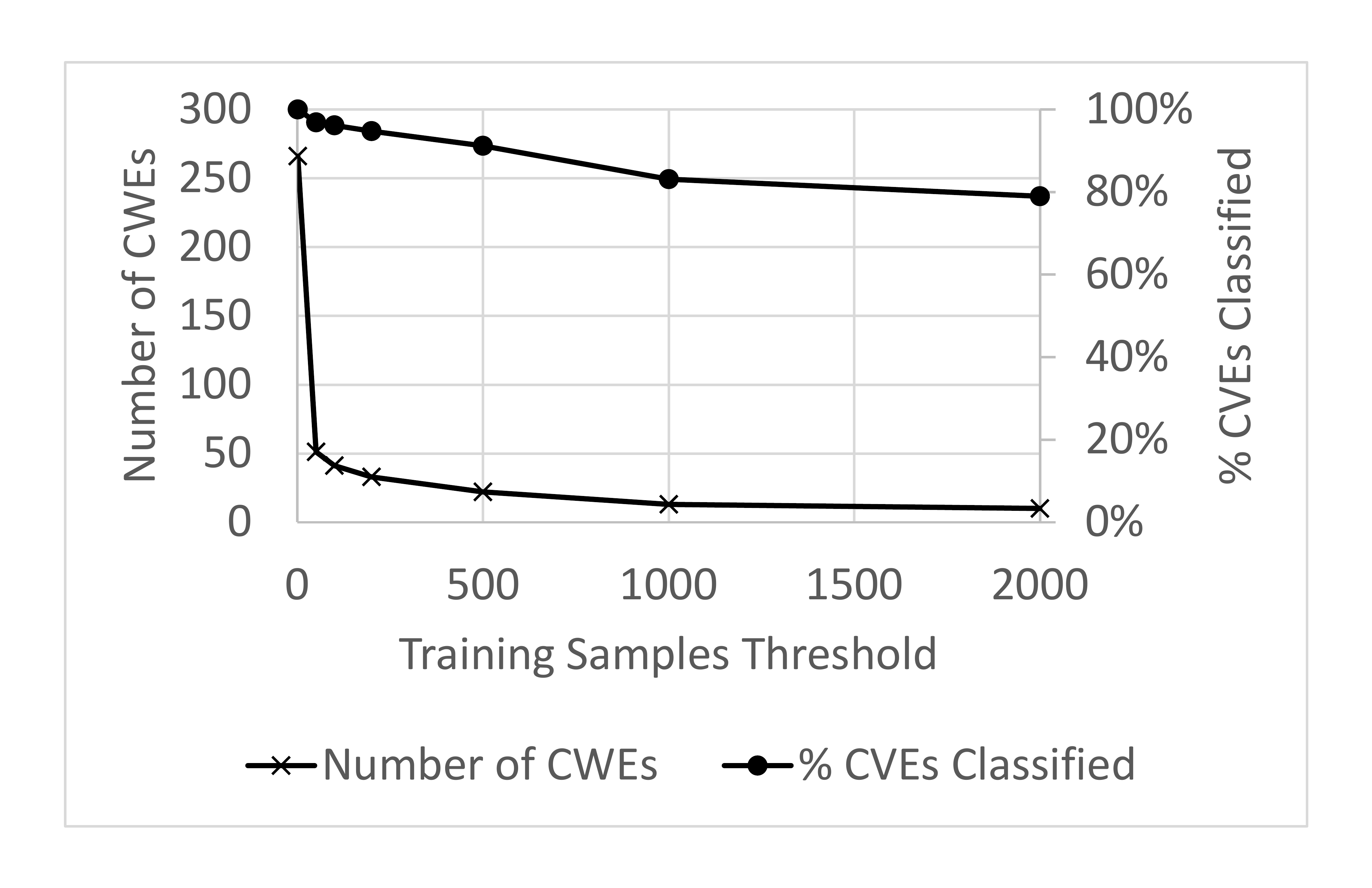}
%   \vspace{-1cm}
  \caption{Coverage of vulnerability to common weakness classification}
  \label{fig:cve-cwe-coverage}
\end{figure}

The accuracy of the results depends on the number of vulnerabilities that we had to train the machine learning algorithms. We define the sample threshold as the minimum number of vulnerabilities that we had available for a specific common weakness during the training process. Setting a sample threshold to $200$ for example, means that we are evaluating our classification procedure and calculating the F-score only for common weaknesses for which there were at least 200 mappings during the training process. Figures~\ref{fig:cve-cwe-fscore}~and~\ref{fig:cve-cwe-coverage} show the classification results and the coverage of the approach. 
% \vspace{-1cm}

Table~\ref{table:cwe-detail} provides a detailed view of the precision, recall and F-score for the $22$ most popular CWEs, for which we had at least 500 vulnerabilities mapped in the training set. Note that the support here refers to the number of samples that we had in the testing set.

To summarize these experiments, there were about $170,000$ vulnerabilities published as of June 2022.  Out of these, around $113,000$ vulnerabilities were mapped to a valid MITRE Common Weakness, the remaining CVEs were not mapped or mapped to an obsolete CWE.  We can use the algorithm to automatically classify these vulnerabilities to CWEs, with a relatively high accuracy, and then use the second step of our workflow to map the vulnerabilities to the MITRE ATT\&CK techniques.

\begin{table*}
  \caption{Classification accuracy for 22 most common weaknesses}
  \label{table:cwe-detail}
  \begin{tabular}{cccccc}
    \toprule
    CWE & Description & Precision & Recall & F-Score & Support\\
    \midrule
    CWE-20 & Improper Input Validation & 1 & 0.86 & 0.93 & 2716 \\
    CWE-22 & Path Traversal & 1 & 0.8 & 0.89 & 865 \\
    CWE-59 & Link Following & 0.99 & 0.76 & 0.86 & 203 \\
    CWE-78 & OS Command Injection & 0.74 & 0.61 & 0.67 & 600 \\
    CWE-79 & Cross-site Scripting & 1 & 0.96 & 0.98 & 3449 \\
    CWE-89 & SQL Injection & 0.96 & 0.94 & 0.95 & 1627 \\
    CWE-94 & Code Injection & 1 & 0.78 & 0.87 & 2200 \\
    CWE-98 & PHP Remote File Inclusion & 0.89 & 0.95 & 0.92 & 361 \\
    CWE-119 & Buffer Overflow & 1 & 0.93 & 0.96 & 2153 \\
    CWE-121 & Stack-based Buffer Overflow & 0.26 & 0.62 & 0.37 & 173 \\
    CWE-122 & Heap-based Buffer Overflow & 0.21 & 0.85 & 0.33 & 279 \\
    CWE-125 & Out-of-bounds Read & 0.99 & 0.53 & 0.69 & 322 \\
    CWE-190 & Integer Overflow or Wraparound & 0.98 & 0.57 & 0.72 & 207 \\
    CWE-200 & Exposure of Sensitive Information & 0.99 & 0.78 & 0.87 & 2348 \\
    CWE-287 & Improper Authentication & 0.99 & 0.57 & 0.73 & 2171 \\
    CWE-352 & Cross-Site Request Forgery (CSRF) & 0.92 & 0.78 & 0.85 & 541 \\
    CWE-416 & Use After Free & 0.47 & 0.7 & 0.56 & 303 \\
    CWE-434 & Unrestricted File Upload & 0.9 & 0.55 & 0.68 & 250 \\
    CWE-451 & UI Misrepresentation of Critical Info & 0.67 & 0.38 & 0.48 & 197 \\
    CWE-476 & NULL Pointer Dereference & 0.92 & 0.72 & 0.81 & 203 \\
    CWE-618 & Exposed Unsafe ActiveX Method & 0.98 & 0.77 & 0.86 & 236 \\
    CWE-787 & Out-of-bounds Write & 0.91 & 0.77 & 0.84 & 166 \\
    \bottomrule
  \end{tabular}
\end{table*}

\subsection{Similarity Search}
We have used similarity search to associate vulnerabilities reported in the NVD to one or several techniques defined in the MITRE ATT\&CK framework. We have created embeddings for the vulnerabilities and the attack techniques using several models, and we used cosine similarity for those embeddings to find the top five MITRE ATT\&CK techniques for each vulnerability.  We have got the best similarity search results by using Doc2Vec~\cite{DBLP:journals/corr/LeM14}, that we briefly describe below.  Doc2Vec provides a numeric representation of text documents.  Doc2Vec is heavily based on Word2Vec \cite{mikolov2013efficient}, a technique used to generate representation vectors out of words, and capture syntactic and semantic word similarities.  Word2Vec relies on two models to capture word similarities.  Continuous bag of words predicts the middle word based on surrounding context words.  Skip gram predicts words within a window before and after the current word.  To generate the numerical representation of a document, Doc2Vec adds another vector (document id) to the Word2Vec model.  Instead of using only words for the continuous bag of words and skip gram models, Doc2Vec also adds a document unique feature vector.  When training the word vectors for a document, the document vector is trained as well, and in the end of training, it holds a numeric representation of the document.  

\subsubsection{Evaluation Metric}
We have evaluated the accuracy of the similarity search using the Mean Reciprocal Rank (MRR). For a single query, the reciprocal rank is \(\frac{1}{rank}\) where \emph{rank} is the position of the highest-ranked correct answer.  If no correct answer was returned, the reciprocal rank is $0$.

\subsubsection{Models Used to Generate Embeddings}
We have experimented with several models to create the vulnerability and technique embeddings.  First, we have built and trained Doc2Vec models. We have preprocessed the technique and vulnerability descriptions by removing stop words, converting to lower case, clearing punctuation, and shortening words to their stems.  Next, we separated technique descriptions into words and assigned a label to each description.  Finally, we created the Doc2Vec model, built the vocabulary and trained the model.  After fine-tuning the Doc2Vec model parameters, we found out that the best results are obtained with 60 training epochs, a vector size of 120, a minimum word count of 9. We also found out that the Doc2Vec model returns the best results when we are training it using only the MITRE ATT\&CK techniques descriptions.  Adding the vulnerability descriptions to the training set has substantially lowered the MRR results, by an order of magnitude.  We think this is due to the differences in vocabulary used to describe the vulnerabilities and the attack techniques.  Next, we have experimented with several pre-trained BERT models from \cite{huggingface}. Finally, we used Spacy to split technique and vulnerability descriptions into sentences (using the ‘en\_core\_web\_lg’ model), and the Google Universal Sentence Encoder to generate the embeddings for the similarity search. 

\subsubsection{Ground Truth Data Set}
To evaluate the accuracy of the similarity search models, we have built a ground-truth set of mappings between vulnerabilities and attack techniques.  We have identified from the MITRE ATT\&CK website all the vulnerabilities that are referenced in the \emph{Procedure Examples} section of the attack technique descriptions.  For example, the \emph{Procedure Examples} section of the technique \emph{T1211 - Exploitation for Defense Evasion}, mentions that \emph{"APT28 has used CVE-2015-4902 to bypass security features"}, and therefore provides a mapping of the vulnerability \emph{CVE-2015-4902} to attack technique \emph{T1211}.  The ground truth set contains 46 unique vulnerabilities retrieved by parsing the attack technique descriptions, as described above. Additionally, we have manually mapped these vulnerabilities to more attack techniques.  We have also manually mapped to the MITRE ATT\&CK techniques $100$ other vulnerabilities, bringing the size of our ground truth data set to $146$ vulnerabilities.

\subsubsection{Similarity Search Results}

Table~\ref{table:sim-search-results} shows the evaluation results for the various similarity search models. We have added to this table the MRR that we have obtained by performing a multi-label hierarchical classification of the vulnerabilities in the ground truth set to CWEs, and then mapping the CWEs to attack techniques.  This MRR measurement assumes that all the vulnerabilities are mapped to CWEs.  In reality, $70\%$ of the vulnerabilities are already mapped to CWEs, so we get an MRR of $1$ for those vulnerabilities.  The remaining $30\%$ are mapped with an MRR of $0.823$, which leaves us with a final MRR of $0.947$.

\begin{table}
  \caption{Mapping of Mitre ATT\&CK techniques to vulnerabilities}
  \label{table:sim-search-results}
  \begin{tabular}{ cc }
      \toprule
      Similarity Search Model & MRR \\
      \midrule
      \makecell{Lookup of Existing Mappings and\\Multi-label Classification using CWEs} & 0.947\\
      Multi-label Classification using CWEs & 0.823\\
      Doc2Vec & 0.465\\
      google/universal-sentence-encoder + Spacy & 0.147\\
      bert-large-cased-whole-word-masking & 0.105\\
      bert-base-uncased & 0.070\\
      bert-large-uncased & 0.067\\
      bert-base-cased & 0.054\\
      albert-base-v1 & 0.050\\
      roberta-base   & 0.040\\
      \bottomrule
  \end{tabular}
\end{table}
% \scriptsize
% \begin{table}
%   \centering
%   \caption{Mapping of Mitre ATT\&CK techniques to vulnerabilities}
%   \begin{tabular}{ cc }
%       \hline\noalign{\smallskip} \textbf{Similarity Search Model} & \textbf{MRR}\\
%       \noalign{\smallskip}\hline
%       Multi-label Classification using CWEs & 0.823\\
%       Doc2Vec & 0.465\\
%       google/universal-sentence-encoder + Spacy & 0.147\\
%       bert-large-cased-whole-word-masking & 0.105\\
%       bert-base-uncased & 0.070\\
%       bert-large-uncased & 0.067\\
%       bert-base-cased & 0.054\\
%       albert-base-v1 & 0.050\\
%       roberta-base   & 0.040\\
%       \noalign{\smallskip}\hline
%   \end{tabular}
%   \label{table:sim-search-results}
% \end{table}
% \normalsize
\subsection{Discussion}
As shown in Table~\ref{table:sim-search-results}, mapping vulnerabilities to attack techniques using CWEs has an accuracy that is $2.04$ times better than that of the best similarity search model (using a Doc2Vec model that is trained on the descriptions of the attack techniques).  As shown in Figures~\ref{fig:cve-cwe-fscore}~and~\ref{fig:cve-cwe-coverage}, we can map around $91\%$ of all available vulnerabilities to the 22 most common weaknesses (for which we have at least 500 samples in our training set), with an average micro and macro F-Score of $0.81$.  So at the cost of maintaining a table of mapping of CWEs to attack techniques, for 22 common weaknesses we get an accurate classification to attack techniques for about 91\% of available vulnerabilities.  If we drop the sample threshold to $200$, we cover $95\%$ of the existing vulnerabilities through $33$ common weaknesses with an average (micro and macro) F-Score of $0.74$.  By further dropping the sampling threshold to $100$, we cover $41$ common weaknesses and $96\%$ of the vulnerabilities with an average F-Score of $0.7$, and for a sampling threshold of $50$, we cover $51$ common weaknesses and $97\%$ of the vulnerabilities with an average F-Score of $0.65$.  The remaining $3\%$ of vulnerabilities are mapped to an additional $215$ common weaknesses, and, because of a lack of training samples, we cannot perform an accurate classification for these common weaknesses, hence the macro F-Score drops to $0.13$.

\section{Deployment}
\label{sec:deployment}

\begin{table*}
  \caption{Analysis of the log4j vulnerabilities}
  \label{table:log4j-results}
  \begin{tabular}{ cccccc }
      \toprule
      Vulnerability & \makecell{CVSS V3\\Score} & \makecell{POC\\Exploits} & CWEs & \makecell{Attack\\Techniques} & \makecell{Threat\\Actors} \\
      \midrule
      CVE-2021-44228 & 10 & 416 & \makecell{20: Improper Input Validation\\400: Uncontrolled Resource Consumption\\ 502: Deserialization of Untrusted Data} & 15 & 50 \\
      CVE-2021-44832 & 6.6 & 3 & \makecell{20: Improper Input Validation\\74: Injection}& 9 & 37 \\
      CVE-2021-45046 & 9 & 13 & 502: Deserialization of Untrusted Data & 5 & 18 \\
      CVE-2021-4104 & 7.5 & 2 & 502: Deserialization of Untrusted Data & 5 & 18 \\
      CVE-2021-44530 & 9.8 & 0 & \makecell{20: Improper Input Validation\\74: Injection} & 9 & 37 \\
      CVE-2021-45105 & 5.9 & 10 & \makecell{20: Improper Input Validation\\674: Uncontrolled Recursion} & 9 & 37 \\
      CVE-2022-21704 & 5.5 & 1 & 276: Incorrect Default Permissions & 29 & 62 \\
      CVE-2022-23302 & 8.8 & 1 & 502: Deserialization of Untrusted Data & 5 & 18 \\
      CVE-2022-23305 & 9.8 & 3 & 89: SQL Injection & 4 & 14 \\
      CVE-2022-23307 & 9.8 & 1 & 502: Deserialization of Untrusted Data & 5 & 18 \\
      \bottomrule
  \end{tabular}
\end{table*}
We have implemented and deployed two services.  First, we are using a machine learning model to map vulnerabilities to common weaknesses.  We need this service because not all vulnerabilities are mapped by default to common weaknesses, some vulnerabilities are mapped to deprecated CWEs, and the classification can be refined using machine learning, by mapping the same vulnerability to common weaknesses that are further from the CWE hierarchy root, and therefore more specific.  A security SME can visualize the mappings of vulnerabilities to common weaknesses returned by the model and the accuracy of the system can be further improved by periodically reviewing the mappings returned by the model, and using Active Learning to incorporate the Security SMEs feedback. 

 Second, we provide an API interface to a service that, given a set of vulnerabilities, returns the attack techniques and the threat actors associated with each vulnerability.  In the next section, we explain in detail how we used this service to analyze the log4j security breach.

\subsection{Analyzing the log4j Security Breach}
The Log4j vulnerabilities~\cite{apache-log4j, log4j-mstic} represent a complex and high-risk situation for companies across the globe. The log4j open-source component is widely used across many suppliers' software and services.  The remote code execution (RCE) vulnerabilities in Apache Log4j referred to as "Log4Shell" (CVE-2021-44228, CVE-2021-45046, CVE-2021-44832) has presented a new attack vector and gained broad attention due to its severity and potential for widespread exploitation.  When successfully exploited, these vulnerabilities could allow an attacker who can control log messages or log message parameters to execute arbitrary code loaded from LDAP servers when message lookup substitution is enabled.  By nature of Log4j being a component, the vulnerabilities affect not only applications that use vulnerable libraries, but also any services that use these applications, so customers may not readily know how widespread the issue is in their environment.

Immediately after the disclosure of the first three RCE vulnerabilities,~\cite{log4j-mstic} warned of nation-state actors attempting to exploit the Log4Shell vulnerability in Log4j, and in the successive days, other flaws (CVE-2021-45105, CVE-2021-44530, CVE-2021-4104, CVE-2022-23305, CVE-2022-23302, CVE-2022-23307, and CVE-2022-21704) were discovered in the library that threat actors attempted to exploit in the wild.

Table~\ref{table:log4j-results} shows a summary of our analysis of the log4j vulnerabilities.  To place our analysis in a broader context, we have added to the table two columns, derived from other sources: the \textit{CVSS V3 Score} and the \textit{POC Exploits}.  The Common Vulnerability Scoring System (CVSS) is an open framework for communicating the characteristics and severity of software vulnerabilities along several dimensions.  The assessment is summarized by assigning each vulnerability a score ranging from 0 to 10.  The POC Exploits column counts, for each vulnerability, the number of proof-of-concept scripts and exploits that are released from security researchers around the world in repositories like ExploitDB~\cite{exploitdb} or Rapid7 Metasploit~\cite{metasploit}.  A larger number of POC Exploits for a given vulnerability signals research interest for that vulnerability, which potentially can be correlated with a larger number of attacks.  As shown in Table~\ref{table:log4j-results}, the CVSS Score and the number of POC exploits can diverge for certain vulnerabilities.  CVE-2021-44530 is rated as critical (9.8) by CVSS, but it has no exploits associated with it.  On the other side, CVE-2021-45105 has a medium CVSS score (5.9), but has 10 exploits associated with it.

The CVSS score and the number of POC exploits associated with a vulnerability tell us how susceptible a vulnerability is to be exploited, and how many exploits have been published for a given vulnerability, but they do not tell us what threat actors are most likely to explore a specific vulnerability.  Known Threat Actors, as defined and tracked within MITRE ATT\&CK, are sophisticated adversaries that often focus on specific geographic regions or industries, and have specific motivations: political, espionage, and financial gains are the most prevalent.  Depending on its industry, and geographic location, a corporation might be paying particular attention to a specific set of threat actors.  Returning a list of threat actors susceptible to exploit a vulnerability allows customers to focus on vulnerabilities with a lower CVSS score, or a lower number of POC exploits.  

To illustrate the way in which our analysis works, we will follow the process of mapping the original log4j vulnerability CVE-2021-44228 to attack techniques and threat actors.  Vulnerability CVE-2021-44228 is mapped in the National Vulnerability Database (NVD) to three Common Weaknesses: CWE-20: Improper Input Validation, CWE-400: Uncontrolled Resource Consumption, and CWE-502: Deserialization of Untrusted Data. 

CWE-20 maps to nine ATT\&CK techniques: T1055 Process Injection, T1203 Exploitation for Client Execution, T1211 Exploitation for Defense Evasion, T1554 Compromise Client Software Binary, T1559.001 Component Object Model, T1562.003 Impair Command History Logging, T1565 Data Manipulation, T1574.006 Dynamic Linker Hijacking, and T1574.007 Path Interception by PATH Environment Variable. CWE-400 maps to one ATT\&CK technique: T1499 Endpoint Denial of Service  CWE-502 maps to five ATT\&CK techniques: T1059 Command and Scripting Interpreter, T1134.002 Create Process with Token, T1134.001 Token Impersonation/Theft, T1550.004 Web Session Cookie, and T1134 Access Token Manipulation.  

There are $50$ threat actors associated with the attack techniques enumerated above.  To check the accuracy of this prediction, we looked at a log4j research report from the Microsoft 365 Defender Threat Intelligence Team~\cite{log4j-mstic}, where CVE-2021-44228 is reported as being used "by multiple tracked nation-state activity groups originating from China, Iran, North Korea, and Turkey".  In particular, the Phosphorus and Hafnium threat actor groups are singled out as actively using this vulnerability.  Looking at the list of threat actors returned by our analysis, we can find Phosphorus (under the alias Magic Hound, and associated with the attack technique T1059), and Hafnium (associated with the attack technique T1203).  Moreover, the list of threat actors returned by our analysis includes other nation-state threat actors, such as Russian groups APT28 (associated with the attack techniques T1203, T1134.001, T1211) and APT29 (linked to T1550.004, and T1203), North Korean groups APT37 (linked to T1055 and T1203) and Lazarus Group (linked to T1134.002 and T1203), Chinese groups Hafnium (linked to T1203) and APT41/Wicked Panda (linked to T1055 and T1203), Iranian groups Phosphorus (linked to T1059), Muddy Water (linked to T1559.001 and T1203) and Fox Kitten (linked to T1059).  We can see a significant overlap between the activity reported in~\cite{log4j-mstic} and the prediction from our analysis.

\subsection{Use Cases}
\label{subsec:use_cases}
The vulnerabilities mappings to threat actors and attack techniques are stored in a database that is accessible through an API to other services, such as risk and vulnerability management, penetration testing, or security event management.

\subsubsection{Risk Prioritization}
One area that the mapping of CVEs to MITRE ATT\&CK techniques can be useful is vulnerability prioritization. There are usually certain attack techniques that threat actors use in order for exploitation. By mapping CVEs to attack techniques and later to threat actors, one can be on alert on CVEs that are exploited by threat actors that specifically target the industry that the business is operating.

\subsubsection{Penetration Testing and Adversary Simulation}
Another area where the mapping of CVEs to MITRE ATT\&CK techniques can be useful is penetration testing and adversary simulations performed by red teams. By gaining the knowledge of attackers and their actual exploits for CVEs that are specific to the target industry, red teams can design their simulation attack and penetration testing strategies to those specific vulnerabilities and their exploits to better be prepared for attacks. 

\subsubsection{Security Event Management}
Finally, an area where mapping of CVEs to MITRE ATT\&CK techniques can be useful is security incident and event management. One can correlate the events that are mapped to MITRE ATT\&CK techniques to the existing vulnerabilities that are mapped to the same MITRE ATT\&CK techniques and better organize the blue team to defend the information assets and protect from attackers.
\section{Related Work}
\label{sec:related}

Recent works~\cite{10.1145/3465481.3465758, threatzoom, DBLP:journals/corr/abs-2010-00533} address the problem of mapping vulnerabilities to MITRE ATT\&CK techniques.  In~\cite{10.1145/3465481.3465758}, the authors propose a Multi-Head Joint Embedding Neural Network model to automatically map vulnerabilities to ATT\&CK techniques.  They develop a new unsupervised labeling technique to address the problem of lack of labels for this task, where they enrich the vulnerabilities with a curated knowledge base of 50 mitigation strategies, which help the model to learn both attacker and defender view of a given vulnerability.  An inherent limitation of this approach is that the model learns from an incomplete knowledge base. Furthermore, the method generates the ground truth data set against which it is evaluated, and hence does not provide a strong evidence that the CVE to MITRE ATT\&CK accuracy is reliable for comparison purpose.  Another limitation of this approach is its reliance on a publicly unavailable data set, and lack of explainability for the generated mappings. In~\cite{threatzoom}, the authors propose a methodology for mapping vulnerabilities to common weaknesses.  While we were not able to reproduce their results, in terms of accuracy of the prediction, we have utilized similar techniques in the mapping of vulnerabilities to common weaknesses.  We have built up upon the work in~(\cite{threatzoom} to classify vulnerabilities to CWEs, and apply this mapping to provide, through mapping to attack techniques, additional intelligence about the vulnerabilities.  In~\cite{DBLP:journals/corr/abs-2010-00533}, the authors link MITRE ATT\&CK Techniques, the  Common Weakness Enumerations (CWE), Common Vulnerabilities and Exposures (CVE), and Common Attack Pattern Enumeration and Classification list (CAPEC attacks) in an aggregate data graph called BRON. They analyze the collection of sources to BRON to provide a view of the extent and range of the coverage and blind spots of public data sources, but they do not tackle the problem of mapping vulnerabilities to attack techniques.

Previous work on vulnerability analysis has focused on using machine learning to predict if a vulnerability will be exploited and when a vulnerability will be exploited \cite{chen2019using}. In \cite{chen2019using}, Chen et al. use CVE data maintained by MITRE \cite{cves}, CVE-related Twitter Discussion data obtained from Twitter as described in \cite{sabottke2015vulnerability}, Exploit DB (EDB) data \cite{exploitdb}, and data from Symantec Intrusion Protection Signature \cite{symantec} to predict when a vulnerability will be exploited. Similarly, \cite{sabottke2015vulnerability} also uses Twitter to predict which vulnerabilities will most likely be exploited. Bozorgi et al. "predict whether and how soon a vulnerability is likely to be exploited" \cite{bozorgi2010beyond} using vulnerability data from the Open Source Vulnerability Database (OSVDB) \cite{osvdb} and the MITRE Common Vulnerabilities and Exposures (CVE) database \cite{cves}. \cite{edkrantz2015predicting} predicts which vulnerabilities will be exploited based on previous exploit patterns using vulnerability data from the National Vulnerabilty Database (NVD) \cite{nvd} and the Exploit Database (EDB) \cite{exploitdb} .

\section{Conclusions and Future Work}
\label{sec:ftr}

In this work, we have built a methodology of mapping vulnerabilities to MITRE ATT\&CK techniques.  This problem is difficult for two reasons: (1) there is very little labelled data available, and (2) we are trying to map two different vocabularies, as vulnerabilities are written by developers, while the attack techniques are defined and used by security professionals and threat actors.  First, we have attempted to resolve the mapping problem by using similarity search.  But the different vocabularies used to describe vulnerabilities and attack techniques are an explanation for the poor accuracy that we have got in the similarity search evaluation.  Our evaluation showed that our approach performs better results by following a two-step classification process - first by mapping vulnerabilities to common weaknesses, and second mapping common weaknesses to attack techniques.  In future work, we are looking to explore in more depth the applications of this classification methodology to vulnerability management, penetration testing, risk management, and management of security events.

\bibliographystyle{ACM-Reference-Format}
\bibliography{bibliography}


\begin{thebibliography}{26}

%%% the Web address.

\ifx \showCODEN    \undefined \def \showCODEN     #1{\unskip}     \fi
\ifx \showDOI      \undefined \def \showDOI       #1{#1}\fi
\ifx \showISBNx    \undefined \def \showISBNx     #1{\unskip}     \fi
\ifx \showISBNxiii \undefined \def \showISBNxiii  #1{\unskip}     \fi
\ifx \showISSN     \undefined \def \showISSN      #1{\unskip}     \fi
\ifx \showLCCN     \undefined \def \showLCCN      #1{\unskip}     \fi
\ifx \shownote     \undefined \def \shownote      #1{#1}          \fi
\ifx \showarticletitle \undefined \def \showarticletitle #1{#1}   \fi
\ifx \showURL      \undefined \def \showURL       {\relax}        \fi
% invisible to TeX
\providecommand\bibfield[2]{#2}
\providecommand\bibinfo[2]{#2}
\providecommand\natexlab[1]{#1}
\providecommand\showeprint[2][]{arXiv:#2}

\bibitem[\protect\citeauthoryear{Aghaei, Shadid, and Al{-}Shaer}{Aghaei
  et~al\mbox{.}}{2020}]%
        {threatzoom}
\bibfield{author}{\bibinfo{person}{Ehsan Aghaei}, \bibinfo{person}{Waseem
  Shadid}, {and} \bibinfo{person}{Ehab Al{-}Shaer}.}
  \bibinfo{year}{2020}\natexlab{}.
\newblock \showarticletitle{ThreatZoom: {CVE2CWE} using Hierarchical Neural
  Network}.
\newblock \bibinfo{journal}{\emph{CoRR}}  \bibinfo{volume}{abs/2009.11501}
  (\bibinfo{year}{2020}).
\newblock
\showeprint[arxiv]{2009.11501}
\urldef\tempurl%
\url{https://arxiv.org/abs/2009.11501}
\showURL{%
\tempurl}


\bibitem[\protect\citeauthoryear{Bozorgi, Saul, Savage, and Voelker}{Bozorgi
  et~al\mbox{.}}{2010}]%
        {bozorgi2010beyond}
\bibfield{author}{\bibinfo{person}{Mehran Bozorgi}, \bibinfo{person}{Lawrence~K
  Saul}, \bibinfo{person}{Stefan Savage}, {and} \bibinfo{person}{Geoffrey~M
  Voelker}.} \bibinfo{year}{2010}\natexlab{}.
\newblock \showarticletitle{Beyond heuristics: learning to classify
  vulnerabilities and predict exploits}. In
  \bibinfo{booktitle}{\emph{Proceedings of the 16th ACM SIGKDD international
  conference on Knowledge discovery and data mining}}.
  \bibinfo{pages}{105--114}.
\newblock


\bibitem[\protect\citeauthoryear{Center}{Center}{2021}]%
        {log4j-mstic}
\bibfield{author}{\bibinfo{person}{Microsoft Threat~Intelligence Center}.}
  \bibinfo{year}{2021}\natexlab{}.
\newblock \bibinfo{title}{Guidance for preventing, detecting, and hunting for
  exploitation of the Log4j 2 vulnerability}.
\newblock
\newblock
\newblock
\shownote{https://www.microsoft.com/security/blog/2021/12/11/guidance-for-preventing-detecting-and-hunting-for-cve-2021-44228-log4j-2-exploitation/.}


\bibitem[\protect\citeauthoryear{Chen, Liu, Park, and Subrahmanian}{Chen
  et~al\mbox{.}}{2019}]%
        {chen2019using}
\bibfield{author}{\bibinfo{person}{Haipeng Chen}, \bibinfo{person}{Rui Liu},
  \bibinfo{person}{Noseong Park}, {and} \bibinfo{person}{VS Subrahmanian}.}
  \bibinfo{year}{2019}\natexlab{}.
\newblock \showarticletitle{Using twitter to predict when vulnerabilities will
  be exploited}. In \bibinfo{booktitle}{\emph{Proceedings of the 25th ACM
  SIGKDD International Conference on Knowledge Discovery \& Data Mining}}.
  \bibinfo{pages}{3143--3152}.
\newblock


\bibitem[\protect\citeauthoryear{Corporation}{Corporation}{2021}]%
        {cwe}
\bibfield{author}{\bibinfo{person}{The~MITRE Corporation}.}
  \bibinfo{year}{2021}\natexlab{}.
\newblock \bibinfo{title}{Common Weakness Enumeration List Version 4.6}.
\newblock
\newblock
\urldef\tempurl%
\url{https://cwe.mitre.org/data/index.html}
\showURL{%
Retrieved February 10, 2022 from \tempurl}


\bibitem[\protect\citeauthoryear{Corporation}{Corporation}{2022a}]%
        {capec}
\bibfield{author}{\bibinfo{person}{The~MITRE Corporation}.}
  \bibinfo{year}{2022}\natexlab{a}.
\newblock \bibinfo{title}{CAPEC: Common Attack Pattern Enumeration and
  Classification}.
\newblock
\newblock
\urldef\tempurl%
\url{https://capec.mitre.org}
\showURL{%
Retrieved February 10, 2022 from \tempurl}


\bibitem[\protect\citeauthoryear{Corporation}{Corporation}{2022b}]%
        {capec-cwe}
\bibfield{author}{\bibinfo{person}{The~MITRE Corporation}.}
  \bibinfo{year}{2022}\natexlab{b}.
\newblock \bibinfo{title}{CWE to CAPEC Mapping}.
\newblock
\newblock
\urldef\tempurl%
\url{https://cwe.mitre.org/data/csv/1000.csv.zip}
\showURL{%
Retrieved February 10, 2022 from \tempurl}


\bibitem[\protect\citeauthoryear{Corporation}{Corporation}{2022c}]%
        {mitre-attack}
\bibfield{author}{\bibinfo{person}{The~MITRE Corporation}.}
  \bibinfo{year}{2022}\natexlab{c}.
\newblock \bibinfo{title}{MITRE ATT\&CK}.
\newblock
\newblock
\urldef\tempurl%
\url{https://attack.mitre.org}
\showURL{%
Retrieved February 10, 2022 from \tempurl}


\bibitem[\protect\citeauthoryear{Corporation}{Corporation}{2022d}]%
        {mitre}
\bibfield{author}{\bibinfo{person}{The~MITRE Corporation}.}
  \bibinfo{year}{2022}\natexlab{d}.
\newblock \bibinfo{title}{MITRE CVE website}.
\newblock
\newblock
\urldef\tempurl%
\url{https://cve.mitre.org}
\showURL{%
Retrieved February 10, 2022 from \tempurl}


\bibitem[\protect\citeauthoryear{Edkrantz and Said}{Edkrantz and Said}{2015}]%
        {edkrantz2015predicting}
\bibfield{author}{\bibinfo{person}{Michel Edkrantz} {and} \bibinfo{person}{Alan
  Said}.} \bibinfo{year}{2015}\natexlab{}.
\newblock \showarticletitle{Predicting Cyber Vulnerability Exploits with
  Machine Learning.}. In \bibinfo{booktitle}{\emph{SCAI}}.
  \bibinfo{pages}{48--57}.
\newblock


\bibitem[\protect\citeauthoryear{Face}{Face}{2022}]%
        {huggingface}
\bibfield{author}{\bibinfo{person}{Hugging Face}.}
  \bibinfo{year}{2022}\natexlab{}.
\newblock \bibinfo{booktitle}{\emph{Models - Hugging Face}}.
\newblock
\urldef\tempurl%
\url{https://huggingface.co/models}
\showURL{%
Retrieved February 10, 2022 from \tempurl}


\bibitem[\protect\citeauthoryear{for Threat-Informed~Defense}{for
  Threat-Informed~Defense}{2021}]%
        {cwe-attack-map}
\bibfield{author}{\bibinfo{person}{The~Center for Threat-Informed~Defense}.}
  \bibinfo{year}{2021}\natexlab{}.
\newblock \bibinfo{title}{Using MITRE ATT\&CK to Describe Vulnerabilities}.
\newblock
\newblock
\newblock
\shownote{\url{https://github.com/center-for-threat-informed-defense/attack_to_cve/blob/master/methodology.md}.}


\bibitem[\protect\citeauthoryear{Foundation}{Foundation}{2021}]%
        {apache-log4j}
\bibfield{author}{\bibinfo{person}{The Apache~Software Foundation}.}
  \bibinfo{year}{2021}\natexlab{}.
\newblock \bibinfo{title}{Apache Log4j Security Vulnerabilities}.
\newblock
\newblock
\urldef\tempurl%
\url{https://logging.apache.org/log4j/2.x/security.html}
\showURL{%
Retrieved February 10, 2022 from \tempurl}


\bibitem[\protect\citeauthoryear{Hemberg, Kelly, Shlapentokh{-}Rothman,
  Reinstadler, Xu, Rutar, and O'Reilly}{Hemberg et~al\mbox{.}}{2020}]%
        {DBLP:journals/corr/abs-2010-00533}
\bibfield{author}{\bibinfo{person}{Erik Hemberg}, \bibinfo{person}{Jonathan
  Kelly}, \bibinfo{person}{Michal Shlapentokh{-}Rothman},
  \bibinfo{person}{Bryn~Marie Reinstadler}, \bibinfo{person}{Katherine Xu},
  \bibinfo{person}{Nick Rutar}, {and} \bibinfo{person}{Una{-}May O'Reilly}.}
  \bibinfo{year}{2020}\natexlab{}.
\newblock \showarticletitle{{BRON} - Linking Attack Tactics, Techniques, and
  Patterns with Defensive Weaknesses, Vulnerabilities and Affected Platform
  Configurations}.
\newblock \bibinfo{journal}{\emph{CoRR}}  \bibinfo{volume}{abs/2010.00533}
  (\bibinfo{year}{2020}).
\newblock
\showeprint[arxiv]{2010.00533}
\urldef\tempurl%
\url{https://arxiv.org/abs/2010.00533}
\showURL{%
\tempurl}


\bibitem[\protect\citeauthoryear{Kuppa, Aouad, and Le-Khac}{Kuppa
  et~al\mbox{.}}{2021}]%
        {10.1145/3465481.3465758}
\bibfield{author}{\bibinfo{person}{Aditya Kuppa}, \bibinfo{person}{Lamine
  Aouad}, {and} \bibinfo{person}{Nhien-An Le-Khac}.}
  \bibinfo{year}{2021}\natexlab{}.
\newblock \showarticletitle{Linking CVE’s to MITRE ATT\&CK Techniques}. In
  \bibinfo{booktitle}{\emph{The 16th International Conference on Availability,
  Reliability and Security}} (Vienna, Austria) \emph{(\bibinfo{series}{ARES
  2021})}. \bibinfo{publisher}{Association for Computing Machinery},
  \bibinfo{address}{New York, NY, USA}, Article \bibinfo{articleno}{21},
  \bibinfo{numpages}{12}~pages.
\newblock
\showISBNx{9781450390514}
\urldef\tempurl%
\url{https://doi.org/10.1145/3465481.3465758}
\showDOI{\tempurl}


\bibitem[\protect\citeauthoryear{Le and Mikolov}{Le and Mikolov}{2014}]%
        {DBLP:journals/corr/LeM14}
\bibfield{author}{\bibinfo{person}{Quoc~V. Le} {and}
  \bibinfo{person}{Tom{\'{a}}s Mikolov}.} \bibinfo{year}{2014}\natexlab{}.
\newblock \showarticletitle{Distributed Representations of Sentences and
  Documents}.
\newblock \bibinfo{journal}{\emph{CoRR}}  \bibinfo{volume}{abs/1405.4053}
  (\bibinfo{year}{2014}).
\newblock
\showeprint[arxiv]{1405.4053}
\urldef\tempurl%
\url{http://arxiv.org/abs/1405.4053}
\showURL{%
\tempurl}


\bibitem[\protect\citeauthoryear{Mikolov, Chen, Corrado, and Dean}{Mikolov
  et~al\mbox{.}}{2013}]%
        {mikolov2013efficient}
\bibfield{author}{\bibinfo{person}{Tomas Mikolov}, \bibinfo{person}{Kai Chen},
  \bibinfo{person}{Greg Corrado}, {and} \bibinfo{person}{Jeffrey Dean}.}
  \bibinfo{year}{2013}\natexlab{}.
\newblock \bibinfo{title}{Efficient Estimation of Word Representations in
  Vector Space}.
\newblock
\newblock
\showeprint[arxiv]{1301.3781}~[cs.CL]


\bibitem[\protect\citeauthoryear{National Institute of Standards and Technology
  (NIST)}{National Institute of Standards and Technology (NIST)}{2015}]%
        {nvd}
National Institute of Standards and Technology (NIST)
  \bibinfo{year}{2015}\natexlab{}.
\newblock \bibinfo{booktitle}{\emph{The National Vulnerability Database
  (NVD)}}.
\newblock National Institute of Standards and Technology (NIST).
\newblock
\newblock
\shownote{\url{https://nvd.nist.gov/}.}


\bibitem[\protect\citeauthoryear{of~Standards and Technology}{of~Standards and
  Technology}{2022}]%
        {nvd-stats}
\bibfield{author}{\bibinfo{person}{National~Institute of Standards} {and}
  \bibinfo{person}{Technology}.} \bibinfo{year}{2022}\natexlab{}.
\newblock \bibinfo{title}{NVD Dashboard}.
\newblock
\newblock
\urldef\tempurl%
\url{https://nvd.nist.gov/general/nvd-dashboard}
\showURL{%
Retrieved February 10, 2022 from \tempurl}


\bibitem[\protect\citeauthoryear{Offensive Security}{Offensive
  Security}{2022}]%
        {exploitdb}
Offensive Security \bibinfo{year}{2022}\natexlab{}.
\newblock \bibinfo{booktitle}{\emph{Offensive Security’s Exploit Database
  Archive}}.
\newblock Offensive Security.
\newblock
\newblock
\shownote{\url{https://www.exploit-db.com/}.}


\bibitem[\protect\citeauthoryear{OSVDB}{OSVDB}{2010}]%
        {osvdb}
OSVDB \bibinfo{year}{2010}\natexlab{}.
\newblock \bibinfo{booktitle}{\emph{The Open Source Vulnerability Database}}.
\newblock OSVDB.
\newblock
\newblock
\shownote{\url{http://osvdb.org/}.}


\bibitem[\protect\citeauthoryear{Rapid7}{Rapid7}{2022}]%
        {metasploit}
Rapid7 \bibinfo{year}{2022}\natexlab{}.
\newblock \bibinfo{booktitle}{\emph{Metasploit - Penetration Testing Software,
  Pen Testing Security}}.
\newblock Rapid7.
\newblock
\newblock
\shownote{\url{https://www.metasploit.com/}.}


\bibitem[\protect\citeauthoryear{Sabottke, Suciu, and Dumitraș}{Sabottke
  et~al\mbox{.}}{2015}]%
        {sabottke2015vulnerability}
\bibfield{author}{\bibinfo{person}{Carl Sabottke}, \bibinfo{person}{Octavian
  Suciu}, {and} \bibinfo{person}{Tudor Dumitraș}.}
  \bibinfo{year}{2015}\natexlab{}.
\newblock \showarticletitle{Vulnerability disclosure in the age of social
  media: Exploiting twitter for predicting real-world exploits}. In
  \bibinfo{booktitle}{\emph{24th $\{$USENIX$\}$ Security Symposium
  ($\{$USENIX$\}$ Security 15)}}. \bibinfo{pages}{1041--1056}.
\newblock


\bibitem[\protect\citeauthoryear{Sokolova and Lapalme}{Sokolova and
  Lapalme}{2009}]%
        {Sokolova:2009:SAP:1542545.1542682}
\bibfield{author}{\bibinfo{person}{Marina Sokolova} {and} \bibinfo{person}{Guy
  Lapalme}.} \bibinfo{year}{2009}\natexlab{}.
\newblock \showarticletitle{A Systematic Analysis of Performance Measures for
  Classification Tasks}.
\newblock \bibinfo{journal}{\emph{Inf. Process. Manage.}} \bibinfo{volume}{45},
  \bibinfo{number}{4} (\bibinfo{date}{July} \bibinfo{year}{2009}),
  \bibinfo{pages}{427--437}.
\newblock
\showISSN{0306-4573}
\urldef\tempurl%
\url{https://doi.org/10.1016/j.ipm.2009.03.002}
\showDOI{\tempurl}


\bibitem[\protect\citeauthoryear{Symantec}{Symantec}{2018}]%
        {symantec}
Symantec \bibinfo{year}{2018}\natexlab{}.
\newblock \bibinfo{booktitle}{\emph{Symantec A-Z Listing of Threats \& Risks}}.
\newblock Symantec.
\newblock
\newblock
\shownote{\url{https://www.broadcom.com/support/security-center/a-z}.}


\bibitem[\protect\citeauthoryear{The MITRE Corporation}{The MITRE
  Corporation}{2020}]%
        {cves}
The MITRE Corporation \bibinfo{year}{2020}\natexlab{}.
\newblock \bibinfo{booktitle}{\emph{Common Vulnerabilities and Exposures}}.
\newblock The MITRE Corporation.
\newblock
\newblock
\shownote{\url{https://cve.mitre.org/}.}


\end{thebibliography}
\end{document}